\documentclass{article}
\usepackage{spconf}
\usepackage{amsmath}
\usepackage{graphicx}
\usepackage{array}
\usepackage{url}
\usepackage{amsthm}
\usepackage{cite}
\usepackage{amsmath,bm}
\usepackage{color}
\usepackage{amssymb}

\usepackage[colorlinks,linkcolor=black,anchorcolor=black,citecolor=black,urlcolor=black]{hyperref}
\hypersetup{
bookmarksnumbered=true,
pdfstartview=FitH
}

\newcommand\normt[1]{\left\lVert#1\right\rVert_2}

\newcommand\normf[1]{\left\lVert#1\right\rVert_F^2}
\newcommand\normfs[1]{\left\lVert#1\right\rVert_F}
\newcommand\normn[1]{\left\lVert#1\right\rVert_*}
\newcommand\hb[1]{\hat{\bm{#1}}}
\newcommand\wb[1]{\widetilde{\bm{#1}}}
\newcommand\tvec[1]{\text{vec}(#1)}
\newcommand\tdiag[1]{\text{diag}(#1)}

\title{
Super-Resolution Harmonic Retrieval of Non-circular Signals}
\name{Yu Zhang$^{\star}$ \qquad Yue Wang$^{\dagger}$ \qquad Zhi Tian$^{\dagger}$ \qquad Geert Leus$^\ddag$
\qquad Gong Zhang$^\star$\thanks{This work was supported in part by the NSFC grants \#62271255, \#61871218, and \#61801211, the US NSF grants \#1939553, \#2003211, \#2128596, \#2136202, and \#2231209, and the ASPIRE project \#14926.}
\address{$^{\star}$ College of EIE, Nanjing University of Aeronautics and Astronautics, Nanjing, China \\
$^{\dagger}$ Department of ECE, George Mason University, Fairfax, VA, USA\\
$^\ddag$ Faculty of EEMCS, Delft University of Technology, Delft, The Netherlands
}
}

\begin{document}
\ninept
\maketitle

\begin{abstract}
This paper proposes a super-resolution harmonic retrieval method for uncorrelated strictly non-circular signals, 
whose covariance and pseudo-covariance present Toeplitz and Hankel structures, respectively.
Accordingly, the augmented covariance matrix constructed by the covariance and pseudo-covariance matrices is not only low rank but also jointly Toeplitz-Hankel structured. 
To efficiently exploit such a desired structure for high estimation accuracy, we develop a low-rank Toeplitz-Hankel covariance reconstruction (LRTHCR) solution employed over the augmented covariance matrix.
Further, we design a fitting error constraint to flexibly implement the LRTHCR algorithm without knowing the noise statistics.
In addition, performance analysis 
is provided for the proposed LRTHCR in practical settings.
Simulation results reveal that the LRTHCR outperforms the benchmark methods in terms of lower estimation errors. 
\end{abstract}
\section{Introduction}
\label{sec:intro}
Harmonic retrieval of non-circular (NC) signals with multiple measurement vectors (MMV) has attracted attention in recent years. 
Many modulated signals used in communication and radar systems are NC in nature such as BPSK, PAM signals and Pseudo-Noise coded sequences, to name a few \cite{Barbaresco2008Noncircularity}. Conventional harmonic estimation methods have been developed for processing and analyzing NC signals~\cite{Charge2001root,Abeida2006MUSIC,Haardt2004Enhancements,Liu2012Direction,Steinwandt2016Gridless,Teng2021Atomic}, among which NC-MUSIC \cite{Charge2001root,Abeida2006MUSIC} and NC-ESPRIT \cite{Haardt2004Enhancements} are popular subspace algorithms. {\color{black}While these methods enjoy enhanced resolution 
thanks to the enlarged manifold in NC scenarios,}
they usually experience low sample efficiency especially in large-antenna systems \cite{Wang2020Enabling, Wang2020Big}. In practice, their estimation performance degrades dramatically under the compression scenario.

To overcome this problem, compressed sensing (CS) based algorithms have been developed to take advantage of the sparsity of sources 
\cite{Liu2012Direction,Steinwandt2016Gridless,Teng2021Atomic}. Unfortunately, the on-grid CS solution in \cite{Liu2012Direction} suffers from limited resolution accuracy due to the basis mismatch. Although the solutions in \cite{Steinwandt2016Gridless,Teng2021Atomic} are based on gridless atomic norm minimization techniques, the solution in \cite{Steinwandt2016Gridless} can only work for odd-dimensional manifolds, and the method designed in \cite{Teng2021Atomic} ignores 
the underlying structural information of covariance and pseudo-covariance. In contrast, a covariance matrix reconstruction approach (CMRA) is proposed for circular signals \cite{Wu2017Toeplitz}, which stems from 
another type of gridless technique termed as low-rank structured covariance reconstruction (LRSCR) and only deals with the covariance matrix. {\color{black}Thus, existing LRSCR misses to utilize the manifold enlargement for increased resolution.} To the best of our knowledge, there is still a lack of LRSCR-type method for NC signals by jointly capturing the structures of both covariance and pseudo-covariance {\color{black}for high estimation accuracy with high sample efficiency}.

To fill this gap, we aim at a sample-efficient LRSCR-based method for harmonic retrieval of NC signals, by jointly exploiting the structural information of both the covariance and pseudo-covariance matrices at the same time. To this end, we first construct an augmented covariance matrix with the covariance and pseudo-covariance matrices of NC signals, which holds a jointly Toeplitz-Hankel structure in the uncorrelated strictly NC case.
Then, we develop a super-resolution harmonic retrieval solution under the LRSCR framework, called low-rank Toeplitz-Hankel covariance reconstruction (LRTHCR), which concurrently utilizes the low-rankness and desired structure of the augmented covariance matrix by solving a structured optimization problem.
Further, a flexible implementation and the performance guarantee of the proposed LRTHCR method are provided under practical settings.
Simulation results verify the advantage of the proposed LRTHCR.

\textit{Notations:} $a$, $\bm{a}$ and $\bm{A}$ denote a scalar, a vector and a matrix, respectively.
$(\cdot)^T$, $(\cdot)^*$, and $(\cdot)^H$ are the transpose, conjugate, and conjugate transpose of a vector or matrix. 
$\text{diag}(\bm{a})$ generates a diagonal matrix with the diagonal elements constructed from $\bm{a}$ and $\text{blkdiag}([\bm{A}_1,\bm{A}_2])$ returns the block diagonal matrix created by aligning $\bm{A}_1$ and $\bm{A}_2$ along the diagonal direction.
$\text{vec}(\cdot)$ stacks all the columns of a matrix into a vector. 
{\color{black}$\text{subvec}(\bm{A})$ returns a vector of size $N(N+1)/2$, which is subsampled from $\text{vec}(\bm{A})$ by keeping only subdiagonal elements of a square matrix $\bm{A}$ of size $N \times N$.} 
$\bm{I}_a$ is an $a$-size identity matrix and {\color{black}$\bm{I}_\Omega$ is generated from an identity matrix by selecting its rows with indices $\Omega$.}
$\text{T}$ and $\text{H}$ represent hermitian Toeplitz and symmetric Hankel matrices, respectively.
$\otimes$ is the Kronecker product. 
$\text{Rank}(\bm{A})$ and $\text{Tr}(\bm{A})$ denote the rank and the trace of $\bm{A}$, respectively. $\mathbb{E}\{\cdot\}$ denotes expectation.

\section{Signal Model}
\label{sec:model}
Consider the problem of harmonic retrieval from strictly NC signal. The NC signal of interest $\bm{x}(t)\in\mathbb{C}^M$ is a linear mixture of $K$ frequency components in the form of
\begin{equation}\label{eq:source}
\begin{split}
     \bm{x}(t)&=\sum_{i=1}^{K}s_i(t)\bm{a}(f_i)=\sum_{i=1}^{K}s_i'(t)e^{j\phi_i}\bm{a}(f_i),\\
     &=\bm{A}(\bm{f})\bm{s}(t)=\bm{A}\bm{\Phi}\bm{s}'(t)\quad t=1,\dots,L,
\end{split}
\end{equation}
where $s_i(t)=s_i'(t)e^{j\phi_i}$ is the complex exponential of the $i$-th source signal at the $t$-th snapshot with real-valued amplitude $s_i'(t)$, $\bm{f}=[f_1,\dots,f_K]^T$ with $f_i\in(-\frac{1}{2},\frac{1}{2}]$ consists of the digital frequencies of $\bm{x}(t)$, 
and $L$ is the number of snapshots. By strict non-circularity, the phase terms $\phi_i$ of source signals are unchanged for all snapshots $t$, unlike circular signals. The manifold matrix
$\bm{A}=\bm{A}(\bm{f})=[\bm{a}(f_1),\dots,\bm{a}(f_K)]$ is made of Vandermonde-structured steering vectors $\bm{a}(f_i)$ of size $M$: 
\begin{equation}\label{eq:stvector}
  \bm{a}(f_i)=[1,\exp(j2\pi f_i),\dots,\exp(j2\pi(M-1)f_i)]^T.
\end{equation}
Further, $\bm{s}(t)=\bm{\Phi}\bm{s}'(t)=[s_1(t),\dots,s_K(t)]^T$ with $\bm{s}'(t)=[s_1'(t){,\dots,}s_K'(t)]^T$ and $\bm{\Phi}{=}\text{diag}(\bm{\phi}){=}\text{diag}([e^{j\phi_1}{,\dots,}e^{j\phi_K}]^T)$ being a diagonal matrix. 

In many applications, $\bm{x}(t)$ is not observed directly, but through subsampling or linear compression via a measurement matrix $\bm{J}\in\mathbb{C}^{N\times M}$ with $N\leq M$. Inflicted by an additive noise $\bm{n}(t)$, the single measurement vector data $\bm{y}(t)\in\mathbb{C}^N$ is given by
\begin{equation}\label{eq:observe}
  \bm{y}(t)=\bm{J}\bm{x}(t)+\bm{n}(t)=\bm{J}\bm{A}\bm{\Phi}\bm{s}'(t)+\bm{n}(t).
\end{equation}
Then, the covariance and the pseudo-covariance of $\bm{y}(t)$ can be respectively expressed as
\begin{equation}\label{eq:covob}
  \begin{split}
     \bm{R}_y&=\mathbb{E}\{\bm{y}(t)\bm{y}^H(t)\}=\bm{J}\bm{R}_x\bm{J}^H+\bm{R}_n, \\
     \bm{C}_y&=\mathbb{E}\{\bm{y}(t)\bm{y}^T(t)\}=\bm{J}\bm{C}_x\bm{J}^T,
  \end{split}
\end{equation}
where $\bm{R}_x$ is the covariance matrix of $\bm{x}(t)$ given by
\begin{equation}\label{eq:covs}
  \bm{R}_x=\mathbb{E}\{\bm{x}(t)\bm{x}^H(t)\}=\bm{A}\bm{R}_{s}\bm{A}^H=\bm{A}\bm{R}_{s'}\bm{A}^H,
\end{equation}
where $\bm{R}_s=\mathbb{E}\{\bm{s}(t)\bm{s}^H(t)\}$, $\bm{R}_{s'}=\mathbb{E}\{\bm{s}'(t)\bm{s}'^T(t)\}$ and $\bm{R}_s=\bm{R}_{s'}=\text{diag}([r_1{,\dots,}r_K]^T)$ with $r_i=\mathbb{E}\{|s_i(t)|^2\}=\mathbb{E}\{s_i'^2(t)\} \!>\!0$ 
for uncorrelated sources. The pseudo-covariance matrix $\bm{C}_x$ of $\bm{x}(t)$ is formed as
\begin{equation}\label{eq:pesucovs}
  \bm{C}_x=\mathbb{E}\{\bm{x}(t)\bm{x}^T(t)\}=\bm{A}\bm{C}_{s}\bm{A}^T=\bm{A}\bm{\Phi}^2\bm{R}_{s'}\bm{A}^T,
\end{equation}
where $\bm{C}_{s}=\mathbb{E}\{\bm{s}(t)\bm{s}^T(t)\}=\bm{\Phi}\mathbb{E}\{\bm{s}'(t)\bm{s}'^T(t)\}\bm{\Phi}=\bm{\Phi}^2\bm{R}_{s'}$. 

In practice, $\bm{R}_y$ and $\bm{C}_y$ are approximated from finite snapshots as $\hb{R}_y=\frac{1}{L}\sum_{t=1}^{L}\bm{y}(t)\bm{y}^H(t)$ and $\hb{C}_y=\frac{1}{L}\sum_{t=1}^{L}\bm{y}(t)\bm{y}^T(t)$, respectively.
Then, the goal of harmonic retrieval of NC signals in this paper is to recover $\{f_i\}_i$ from $\hat{\bm{R}}_y$ and $\hat{\bm{C}}_y$ jointly. {\color{black}Note that $\bm{C}_x=\bm{0}$ and $\bm{C}_y=\bm{0}$ for circular signals, which is a key difference from NC signals.}

\section{Proposed Method}
\label{sec:proposed}
This section proposes a super-resolution harmonic retrieval method with NC signals. {\color{black}We first construct an augmented covariance matrix with the covariance and pseudo-covariance matrices. Then, we develop an effective method to recover this augmented matrix from compressive measurements by exploiting its inherent structural information and low-rankness.} Finally, harmonic retrieval estimation is conducted over the reconstructed augmented covariance matrix.

\subsection{Low-Rank Toeplitz-Hankel Covariance Reconstruction}
\label{subsec:LRTHCR}
Define a compressed augmented measurement vector 
$\bm{z}(t)$ as 
\begin{equation}\label{eq:comaugvec}
  \bm{z}(t){=}
  \left[\begin{array}{c}
    \bm{y}(t) \\
    \bm{y}^*(t)
  \end{array}\right]{=}
                    \bm{J}'{\left[\begin{array}{c}
                                  \bm{x}(t) \\
                                  \bm{x}^*(t)
                                \end{array}\right]}{+}\left[\begin{array}{c}
                                  \bm{n}(t) \\
                                  \bm{n}^*(t)
                                \end{array}\right]{=}\bm{J}'\bm{x}_a(t){+}\bm{n}'(t),
\end{equation}
where $\bm{n}'(t){=}[\bm{n}^T(t),\bm{n}^H(t)]^T$, $\bm{J}'{=}\text{blkdiag}([\bm{J},\bm{J}^*])\in\mathbb{C}^{2N\times 2M}$ is the augmented measurement matrix and $\bm{x}_a(t){=}[\bm{x}^T(t),\bm{x}^H(t)]^T$ is the augmented signal vector. Then, the compressed augmented covariance matrix $\bm{R}_z$ of $\bm{z}(t)$ can be expressed as
\begin{equation}\label{eq:comaugmat}
  \bm{R}_z{=}\mathbb{E}\{\bm{z}(t)\,\bm{z}^H(t)\}{=}\left[\begin{array}{cc}
             \bm{R}_y & \bm{C}_y \\
             \bm{C}_y^* & \bm{R}_y^*
           \end{array}\right]{=}\bm{J}'\bm{R}_a\bm{J}'^H+\bm{R}_{n'},
\end{equation}
where 
\begin{equation}\label{eq:augmat}
  \bm{R}_a=\mathbb{E}\{\bm{x}_a(t)\,\bm{x}^H_a(t)\}=\left[\begin{array}{cc}
                                     \bm{R}_x  & \bm{C}_x \\
                                     \bm{C}_x^*  & \bm{R}_x^*
                                    \end{array}\right]
\end{equation}
denotes the augmented covariance matrix of $\bm{x}_a(t)$ and $\bm{R}_{n'}$ is the covariance of $\bm{n}'(t)$. 
Note that $\bm{R}_a$ in \eqref{eq:augmat} 
contains all the harmonic information while having an enlarged steering vector of length $2M$, which leads to enhanced resolution. 
Thus, we can retrieve the frequencies $\{f_i\}_i$ via NC-based methods, once $\bm{R}_a$ is obtained. Moreover, the sample compressed augmented covariance $\hb{R}_z$ is constructed by $\hb{R}_y$ and $\hb{C}_y$ as $\hb{R}_z=[\hb{R}_y,\hb{C}_y;\hb{C}_y^*, \hb{R}_y^*]$. Then, the task boils down to obtaining an estimate of $\bm{R}_a$ from $\hb{R}_z$.

Assuming uncorrelated signal sources, \eqref{eq:covs} and \eqref{eq:pesucovs} amount to
\begin{equation}\label{eq:eleinc}
     r_{ij}{=}\sum_{k=1}^{K}r_k e^{j2\pi(i{-}j)f_k},\quad
  c_{ij}{=}\sum_{k=1}^{K}r_k e^{j2\phi_k} e^{j2\pi(i+j-2)f_k},
\end{equation}
where $r_{ij}$ and $c_{ij}$ denote the element of $\bm{R}_x$ and $\bm{C}_x$ in the $i$-th row and $j$-th column, respectively.
From \eqref{eq:eleinc},
it is obvious that $\bm{R}_x$ in \eqref{eq:covs} and $\bm{C}_x$ in \eqref{eq:pesucovs} are a positive semi-definite (PSD) hermitian Toeplitz matrix and a symmetric Hankel matrix, respectively, 
\begin{equation}\label{eq:THrep}
     \bm{R}_x=  \text{T},\quad
     \bm{C}_x=  \text{H}.
\end{equation}

Substituting \eqref{eq:THrep} into \eqref{eq:augmat}, we have
\begin{equation}\label{eq:straugmat}
  \bm{R}_a=\left[\begin{array}{cc}
                                     \text{T}  &\text{H} \\
                                     \text{H}^*  &\text{T}^*
                                    \end{array}\right],
\end{equation}
which is a PSD, hermitian, and jointly Toeplitz-Hankel structured matrix. Accordingly, $\bm{R}_a$ is not only a low-rank matrix with rank $K$, but also holds an underlying Toeplitz-Hankel structure. Expanding the LRSCR theory \cite{Li2016Grid,Wang2019Super,Wang2020Efficient,Zhang2022Efficient}, the augmented covariance $\bm{R}_a$ in \eqref{eq:straugmat} can be recovered by
\begin{equation}\label{eq:lrsmr}
\begin{split}
   \wb{R}_a=&\arg\min_{\bm{R}_a} \tau\text{Rank}(\bm{R}_a)+\frac{1}{2}\normf{\hb{R}_z-\bm{J}'\bm{R}_a\bm{J}'^H}\\
     & ~~~~~\text{s.t.}~~~~~\bm{R}_a\succeq 0,~\bm{R}_a~\text{in}~\eqref{eq:straugmat},
\end{split}
\end{equation}
where the first term imposes low rankness, the second least-squares term minimizes the model fitting error based on \eqref{eq:comaugmat}, and $\tau$ is a regularization parameter to balance the model fitting error and the low rankness.

The noncovexity of the objective function in \eqref{eq:lrsmr} makes it an NP-hard problem. To overcome this issue, we relax $\text{Rank}(\bm{R}_a)$ to its convex cone, aka trace norm, which gives rise to the following low-rank Toeplitz-Hankel covariance reconstruction (LRTHCR) for augmented covariance matrix recovery
\begin{equation}\label{eq:LRTHCR}
\begin{split}
   \wb{R}_a=&\underset{\bm{R}_a}{\arg\min} ~\tau\text{Tr}(\bm{R}_a)+\frac{1}{2}\normf{\hb{R}_z-\bm{J}'\bm{R}_a\bm{J}'^H}\\
     & ~~~~~\text{s.t.}~~~~~\bm{R}_a\succeq 0,~\bm{R}_a~\text{in}~\eqref{eq:straugmat}.
\end{split}
\end{equation}
LRTHCR in
\eqref{eq:LRTHCR} can be solved using off-the-shelf semi-definite programming (SDP) based solvers, such as CVX \cite{Grant2014CVX}.

\subsection{LRTHCR Implementation via Fitting Error Constraint}
\label{subsec:const}
The regularization parameter $\tau$ in \eqref{eq:LRTHCR} is related to the noise statistics, which is difficult to set when such statistics are unavailable in practice. {\color{black}Given sufficient MMV, we design a fitting error constraint to
reformulate the original Lagrangian form of \eqref{eq:LRTHCR} for flexible implementation of the LRTHCR formula,} 
where the pre-fixed parameter for bounding the fitting error can be set without knowing the noise statistics.
{\color{black}To this end, we resort to converting covariance and pseudo-covariance matrices to their vectorized forms to facilitate the design. Consider the noise $\bm{n}(t)$ modeled as circular Gaussian noise with distribution $\text{N}(\bm{0},\text{diag}(\bm{\sigma}^2))$ having $\bm{R}_n{=}\text{diag}(\!\bm{\sigma}^2)$.} Let $\bm{r}_y{=}\tvec{\!\bm{R}_y\!}\!\in\!\mathbb{C}^{N^2}$ and $\bm{c}_y{=}\text{subvec}(\!\bm{C}_y\!)\!\in\!\mathbb{C}^{{N(N+1)}/{2}}$. Then, the definitions of $\bm{R}_y$, $\bm{C}_y$ in \eqref{eq:covob} and $\bm{R}_x$, $\bm{C}_x$ in \eqref{eq:THrep} lead to
\begin{equation}\label{eq:veccovob}
  \begin{split}
     \bm{r}_y= & (\bm{J}^*\otimes\bm{J})\tvec{\text{T}}+\tvec{\tdiag{\bm{\sigma}^2}}, \\
     \bm{c}_y= & \bm{U}(\bm{J}\otimes\bm{J})\tvec{\text{H}},
  \end{split}
\end{equation}
where $\bm{U}\in\mathbb{C}^{N(N+1)/2\times N^2}$ is the selection matrix that satisfies $\text{subvec}(\bm{C}_y)=\bm{U}\tvec{\bm{C}_y}$.
Since the compressed augmented covariance matrix $\bm{R}_z$ is constructed by $\bm{R}_y$ and $\bm{C}_y$, we can construct a vector containing all the information of $\bm{R}_z$ from $\bm{r}_y$ and $\bm{c}_y$ in the form of
\begin{equation}\label{eq:vecinconst}
    \bm{q}_y=[\bm{r}_y^T,\bm{c}_y^T,\bm{c}_y^H]^T.
\end{equation}
Similarly, the sample vector $\hb{q}_y$ is constituted by $\hb{r}_y=\tvec{\hb{R}_y}$ and $\hb{c}_y=\text{subvec}(\hb{C}_y)$ as $\hb{q}_y=[\hb{r}_y^T,\hb{c}_y^T,\hb{c}_y^H]^T$, which also contains all the information of $\hb{R}_z$.

According to the covariance matching estimation technique for NC signals \cite{Delmas2004Asymptotically}\cite{Delmas2009Asymptotic}, it can be shown that when the complex amplitude vector $\bm{s}(t)$ in \eqref{eq:source} is an NC Gaussian random variable with zero mean, the residual error $\hb{q}_y-\bm{q}_y$ follows an asymptotic complex normal ($\text{AsN}_c$) distribution:
\begin{equation}\label{eq:AsNc}
  \hb{q}_y-\bm{q}_y \sim \text{AsN}_c(\bm{0},\bm{R}_q,\bm{C}_q),
\end{equation}
where $\bm{R}_q$ and $\bm{C}_q$ are the covariance and pseudo-covariance matrices of $\hat{\bm{q}}_y - \bm{q}_y$, respectively.

Then, for ease of determining the pre-fixed parameter for a fitting error constraint, we ignore the effect of $\bm{C}_q$ and assume $\hb{q}_y-\bm{q}_y$ to follow an approximate normal ($\text{AN}$) distribution as $\hb{q}_y-\bm{q}_y \sim \text{AN}(\bm{0},\bm{R}_q)$. 
Accordingly, $\hb{R}_q^{-\frac{1}{2}}(\hb{q}_y-\bm{q}_y) \sim \text{AN}(\bm{0},\bm{I}_{_{2N^2+N}})$, where $\hb{R}_q$ is the estimate of $\bm{R}_q$. Thus, $\normf{\hb{R}_q^{-\frac{1}{2}}(\hb{q}_y-\bm{q}_y)}$ follows an approximate Chi-squared ($\text{A}\chi^2$) distribution:
\begin{equation}\label{eq:Achi}
  \normf{\hb{R}_q^{-\frac{1}{2}}(\hb{q}_y-\bm{q}_y)}\sim\text{A}\chi^2(2N^2+N).
\end{equation}
As a result, the following inequality holds with probability $1-p$:
\begin{equation}\label{eq:constraint}
  \normf{\hb{R}_q^{-\frac{1}{2}}(\hb{q}_y-\bm{q}_y)}\leq \eta_p,
\end{equation}
where the parameter $\eta_p$ can be uniquely determined from \eqref{eq:Achi} by the degrees of freedom $2N^2{+}N$ and a pre-fixed small 
deviation probability $p\ll 1$, which is independent of noise variance $\bm{\sigma}^2$.

Adopting \eqref{eq:constraint} into LRTHCR and having $\bm{q}_y$ expressed in \eqref{eq:vecinconst} with \eqref{eq:veccovob}, the least-squares term in the original objective function of \eqref{eq:LRTHCR} can be recast into the constraint. Then,
the proposed LRTHCR can be implemented without knowing the noise statistics as
\begin{equation}\label{eq:LRTHCRconst}
  \begin{split}
   \wb{R}_a=&\underset{\bm{R}_a}{\arg\min}~\text{Tr}(\bm{R}_a)\\
     & ~~~~~\text{s.t.}~~~~~
                                    \bm{R}_a\succeq 0,~\bm{R}_a~\text{in}~\eqref{eq:straugmat}\\
     & \quad\quad\quad~~ \normf{\hb{R}_q^{-\frac{1}{2}}(\hb{q}_y-\bm{q}_y)}\leq \eta\\
     & \quad\quad\quad~~~ \bm{q}_y~\text{in}~\eqref{eq:vecinconst}.
\end{split}
\end{equation}

\subsection{Harmonic Retrieval}
\label{subsec:retrieval}
By solving the proposed LRTHCR in \eqref{eq:LRTHCR} or \eqref{eq:LRTHCRconst}, we obtain the estimate of the augmented covariance matrix $\wb{R}_a$. It is structured and can be expressed as below by plugging \eqref{eq:covs} and \eqref{eq:pesucovs} into \eqref{eq:augmat}
\vspace{-0.01in}
\begin{equation}\label{eq:augdecom}
  \wb{R}_a=\left[\begin{array}{c}
                   \wb{A} \\
                   \wb{A}^*\wb{\Phi}^{-2}
                 \end{array}\right]\wb{R}_{s'}\left[\wb{A}^H,~\wb{\Phi}^2\wb{A}^T\right]=\bm{H}\wb{R}_{s'}\bm{H}^H,
\end{equation}
where $\bm{H}=[\wb{A}^T,(\wb{A}^*\wb{\Phi}^{-2})^T]^T$ with $\wb{A}$ and $\wb{\Phi}$ being the estimates of $\bm{A}$ and $\bm{\Phi}$, respectively. Given $\wb{R}_a$ in \eqref{eq:augdecom}, the NC-based algorithms can be applied for harmonic retrieval, such as, the NC-MUSIC \cite{Abeida2006MUSIC,Charge2001root} and  NC-ESPRIT \cite{Haardt2004Enhancements}. {\color{black} Note that 
$\bm{H}$ equips with enlarged steering vectors of length $2M$ compared with the original $\bm{A}$ whose steering vectors are of length $M$. Hence, the proposed method offers enhanced resolution.}

\section{Performance Guarantee} 
\label{sec:discussion}
This section presents theoretical analysis of the proposed LRTHCR method by providing its performance guarantee under finite MMV and additive white Gaussian noise, where $\bm{n}(t)$ in \eqref{eq:observe} satisfies $\text{N}(\bm{0},\sigma^2\bm{I}_{_N})$.
Adopting $\bm{J}=\bm{I}_\Omega$ as a random selection matrix, we have $\bm{J}'=\bm{I}_{\Omega'}=\text{blkdiag}([\bm{I}_\Omega,\bm{I}_\Omega])$ with $\Omega'=\Omega\bigcup(\Omega+M)$, and $\bm{R}_{\Omega'}=\bm{I}_{\Omega'}\bm{R}_a\bm{I}^T_{\Omega'}$. {\color{black}Suppose that $\Omega'$ is a complete sparse ruler, which is to ensure identifiability of $\bm{R}_a$ from $\bm{R}_{\Omega'}$ \cite{Gupta2019Design}.} We introduce the noise estimate $\hat{\sigma}^2\bm{I}_{_{2N}}$ with $\hat{\sigma}^2$ being the smallest eigenvalue of $\hb{R}_z$ into \eqref{eq:LRTHCR}, which leads to
\begin{equation}\label{eq:LRTHCRwsig}
\begin{split}
   &\arg\min_{\bm{R}_a} \tau\text{Tr}(\bm{R}_a){+}\frac{1}{2}\normf{\hb{R}_z{-}\bm{R}_{\Omega'}{-}\hat{\sigma}^2\bm{I}_{_{2N}}}\\
     & ~~~~~~\text{s.t.}~~~~\bm{R}_a\succeq 0,~\bm{R}_a~\text{in}~\eqref{eq:straugmat}.
\end{split}
\end{equation}
The introduction of the noise estimate $\hat{\sigma}^2\bm{I}_{_{2N}}$ is to deduce that the solution of \eqref{eq:LRTHCRwsig} can asymptotically approach the ground truth, as stated by the following theorem.

\noindent\emph{\textbf{Theorem 1:}} Let $\bm{R}_a^\star$, ${\sigma}^2$ be the ground truth, and $\wb{R}_a$ be the optimal solution to \eqref{eq:LRTHCRwsig}, respectively. Set
  $\tau\geq (C_1\text{Tr}(\bm{R}_{\Omega'}^\star)+C_2\sigma^2)\sqrt{\frac{\ln L}{L}}$
for some constant $C_1$ and $C_2$.
Then, with probability at least $1-5L^{-1}$, the solution $\wb{R}_a$ to \eqref{eq:LRTHCRwsig} satisfies
\begin{equation}\label{eq:bounderror}
  \normfs{\wb{R}_a-\bm{R}_a^\star}\leq \tau\sqrt{M}(8\sqrt{KM}+1).
\end{equation}

\noindent \textit{Proof:} A sketch of proof is provided here due to page limit.

First, letting $\hb{R}_L=\hb{R}_y-\hat{\sigma}^2\bm{I}_{_{2N}}$, according to the results of \cite{Bunea2015sample} and Lemma~1 in \cite{Negahban2012Unified}, we have
\begin{equation}\label{eq:tauinequ}
  \tau{\geq} 2\normfs{\hb{R}_L{-}\bm{R}_{\Omega'}^\star}\geq 2 \normt{\hb{R}_L{-}\bm{R}_{\Omega'}^\star},
\end{equation}
with $\tau$ defined in Theorem 1. 
Then, based on the optimality of $\wb{R}_a$ to \eqref{eq:LRTHCRwsig},
and Lemma~1 in \cite{Negahban2012Unified}, and applying some inequalities including \eqref{eq:tauinequ}, triangle inequality, Cauchy-Schwartz inequality and $\normn{\bm{B}}\leq\sqrt{K}\normfs{\bm{B}}$ for any hermitian matrix $\bm{B}$ with rank $K$ \cite{zhang2017matrix}, we have
\begin{equation}\label{eq:Rasamieq}
  \normf{\wb{R}_{\Omega'}{-}\bm{R}_{\Omega'}^\star}\leq  8\tau\sqrt{K}\normfs{\wb{R}_a{-}\bm{R}_a^\star}\!+\tau \normfs{\wb{R}_{\Omega'}{-}\bm{R}_{\Omega'}^\star}.
\end{equation}
In addition, based on the property of complete sparse rulers, we have
\begin{equation}\label{eq:sparseruler}
    \normfs{\wb{R}_a-\bm{R}_a^\star} \leq \sqrt{M} \normfs{\wb{R}_{\Omega'}-\bm{R}_{\Omega'}^\star}.
\end{equation}
Finally, substituting \eqref{eq:sparseruler} into \eqref{eq:Rasamieq}, we can conclude the proof. $\hfill\square$

\noindent \textit{Remark 1:} Note that in Theorem 1 the observation set $\Omega'$ is deterministic for a given $\bm{J}=\bm{I}_{{\Omega}}$, and that $8\sqrt{KM}$ is much larger than $1$. Our algorithm yields reliable estimate of $\bm{R}_a$ as long as $L$ is on the order of $(\text{Tr}(\bm{R}^{\star}_{\Omega'})+\sigma^2)^2M^2K$. The MSE in \eqref{eq:bounderror} diminishes as $L$ increases, and as $L\rightarrow \infty$, the ground truth $\bm{R}_a^\star$ can be exactly recovered, which means that the estimate $\wb{R}_a$ of the proposed LRTHCR is statistically consistent in $L$. Moreover, since the frequency estimates can be uniquely determined by the optimal solution of our 
LRTHCR, the estimates $\{\widetilde{f}_i\}_i$ are statistically consistent in $L$.

%

\begin{figure}[!tbp]
\centering
\includegraphics[width=2.35in]{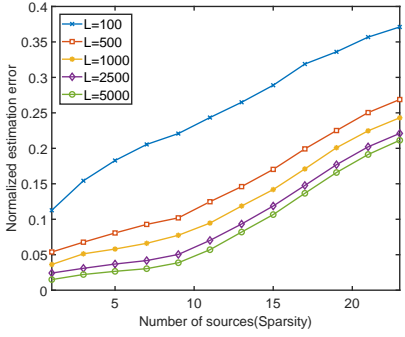}\vspace{-0.1in}
\caption{The normalized estimation error with respect to the number of sources (sparsity level) $K$ for various the number of measurement vectors $L$ for the proposed LRTHCR method with $M{=}13$, $N{=}8$, $\text{SNR}{=}10\text{dB}$ and $M_t{=}300$.} \vspace{-0.12in}
\label{fig:NEE}
\end{figure}

\section{Numerical Simulations}
\label{sec:simulations}
This section presents numerical results to evaluate the performance of the proposed LRTHCR solution. Existing methods such as the conventional NC-MUSIC \cite{Abeida2006MUSIC} and the CMRA (equivalent to the LRSCR) \cite{Wu2017Toeplitz} with the MUSIC algorithm are run for frequency retrieval as benchmarks, while the Cramer-Rao bound (CRB) under full observation is also provided \cite{Delmas2004Stochastic}. {\color{black}The NC-MUSIC is employed for the proposed LRTHCR method for harmonic retrieval.} 
The peak searching step is set to be $10^{-4}$ when the peak searching operator is needed. The root mean squared error (RMSE) is used to measure 
the estimation accuracy of harmonic retrieval as $\text{RMSE}=\frac{1}{K}\sum_{k=1}^{K}\left(\frac{1}{M_t}\sum_{n=1}^{M_t}(\widetilde{f}_k^n-f_k)^2\right)^{\frac{1}{2}}$, where $M_t$ and $\widetilde{f}_k^n$ denote the number of the Monte-Carlo trials and the estimates of $f_k$ in the $n$-th trial, respectively.

First, we conduct an 
experiment to test 
the proposed LRTHCR 
by evaluating 
the influence of the number of measurement vectors $L$ on the recovery of 
the augmented covariance matrix $\bm{R}_a$ in terms of the average normalized estimation error defined as $\mathbb{E}\left\{\normfs{\wb{R}_a-\bm{R}_a^\star}\right\}/\mathbb{E}\left\{\normfs{\bm{R}_a^\star}\right\}$. The frequencies $\{f_i\}_i$ are selected uniformly from $(-\frac{1}{2},\frac{1}{2}]$, and each source signal $s_i(t)$ is generated with $\phi_i$ being selected uniformly from $(0,\pi]$ and $s_i'(t)$ randomly drawn from $\text{N}(0,1)$. As shown in Fig.~\ref{fig:NEE}, the average normalized estimation error decreases for a fixed sparsity level $K$ as $L$ increases, and decreases for a fixed $L$ as $K$ decreases, which verifies the result of Theorem~1. 

\begin{figure}[!tbp]
\centering
\includegraphics[width=2.35in]{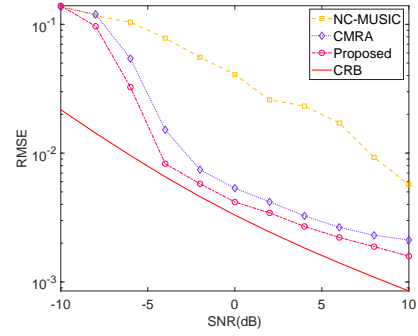}\vspace{-0.1in}
\caption{RMSE versus SNR for the proposed LRTHCR, CMRA, NC-MUSIC and CRB with $M{=}7$, $N{=}4$, $K{=}4$, $L{=}300$ and $M_t{=}300$.}\vspace{-0.12in}
\label{fig:RMSEvsSNR}
\end{figure}

\begin{figure}[!tbp]
\centering
\includegraphics[width=2.35in]{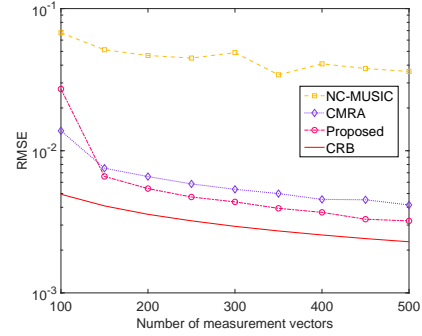}\vspace{-0.1in}
\caption{RMSE versus number of measurement vectors $L$ for the proposed LRTHCR, CMRA, NC-MUSIC and CRB with $M{=}7$, $N{=}4$, $K{=}4$, $\text{SNR}{=}0\text{dB}$ and $M_t{=}300$.}\vspace{-0.12in}
\label{fig:RMSEvsSnap}
\end{figure}

Next, we evaluate the estimation performance of the proposed LRTHCR for harmonic retrieval. 
We consider $\bm{J}=\bm{I}_\Omega$ with selection indices $\Omega=\{1,2,5,7\}$ and there are four equal-power uncorrelated source signals with $\bm{f}=[-0.3,0,0.2,0.4]$ and $\bm{\phi}$ being selected uniformly from $(0,\pi]$. To implement the proposed LRTHCR, $\eta_p$ is determined with $p=0.01$. 
We compare the RMSE of different 
algorithms versus the SNR with $L=300$. Fig.~\ref{fig:RMSEvsSNR} indicates 
that the conventional NC-MUSIC cannot achieve desired performance under the compression scenario, whereas the proposed LRTHCR 
achieves the best performance among these methods. 
Fig.~\ref{fig:RMSEvsSnap} presents the RMSE performance of these approaches for different $L$,  
which reveals a similar trend as that shown in Fig.~\ref{fig:RMSEvsSNR}.

\section{Conclusion}
\label{sec:conclusion}
This paper develops a super-resolution harmonic retrieval method for NC signals. 
An augmented covariance matrix is formed with covariance and pseudo-covariance matrices, which presents an important Toeplitz-Hankel structure for the uncorrelated strictly NC source signals. To fully utilize this desired structural information, a LRTHCR solution is proposed 
and employed for augmented covariance matrix reconstruction from compressive measurements, followed by NC-based subspace algorithms for harmonic retrieval. 
Simulation results validate the merit of the LRTHCR method with higher estimation performance beyond existing benchmarks.

\vfill\pagebreak

\bibliographystyle{IEEEbib}

\end{document}